\documentclass[runningheads]{llncs}
\usepackage[T1]{fontenc}
\usepackage{graphicx}
\usepackage{amsmath,amssymb,amsfonts}
\usepackage{booktabs}
\usepackage{xurl}
\usepackage[colorlinks=true,linkcolor=black,citecolor=black,urlcolor=blue]{hyperref}

\usepackage{etoolbox}

\makeatletter
\patchcmd{\maketitle}
  {\else
      \newpage
      \global\@topnum\z@}
  {\else
      \global\@topnum\z@}
  {}{}

\patchcmd{\@maketitle}{\vskip .8cm}{\vskip .35cm}{}{}

\patchcmd{\@maketitle}{\@author\vskip.35cm}{\@author\vskip.15cm}{}{}
\makeatother

\begin{document}
\title{Quantization-Aware Neuromorphic Architecture for Skin Lesion Classification on Resource-Constrained Devices}
\titlerunning{Quantization-Aware Neuromorphic Architecture}
%
%
\author{%
Haitian Wang\inst{1,3}\thanks{Corresponding author: \email{haitian.wang@curtin.edu.au}} \and
Xia Cheng\inst{2,3} \and
Xinyu Wang\inst{3} \and
Fiona Wei\inst{3} \and
Zichen Geng\inst{3}}
\authorrunning{H. Wang et al.}

\institute{%
School of Electrical Engineering, Computing and Mathematical Sciences,\\
Curtin University, Perth, WA, Australia
\and
School of Information Technology,
Murdoch University, Perth, WA, Australia
\and
Department of Computer Science and Software Engineering,\\
The University of Western Australia, Perth, WA, Australia
}

\maketitle              

\vspace{-9mm}
\begin{abstract}
On-device skin lesion analysis is constrained by the compute and energy cost of conventional CNN inference and by the need for lightweight calibration under clinical data shift. Neuromorphic processors provide event-driven sparse computation, but practical deployment is often limited by CNN-to-SNN conversion failures, including unsupported operators, quantization distortion, and accuracy degradation under class imbalance. We propose QANA, a quantization-aware CNN backbone within an end-to-end pipeline for conversion-stable neuromorphic execution. QANA improves conversion robustness by bounding intermediate activations, aligning normalization with low-bit quantization, and replacing conversion-fragile components with spike-compatible transformations. Efficient representation is achieved through Ghost-based feature generation, spatially-aware efficient channel attention, and squeeze-and-excitation modules whose operations can be quantized, folded, or lowered into integer graph operations. The quantized projection head produces SNN-ready logits and supports lightweight readout calibration without full retraining or data offloading. On HAM10000, QANA achieves 91.6\% Top-1 accuracy and 91.0\% macro F1, improving the strongest converted SNN baseline by 3.5 points in accuracy and 12.1 points in macro F1. On a clinical dataset, QANA achieves 90.8\% Top-1 accuracy and 81.7\% macro F1, improving the strongest baseline by 3.2 points in accuracy and 3.6 points in macro F1. On BrainChip Akida, QANA runs in 1.5\,ms with 1.7\,mJ per image, outperforming the strongest size-matched Akida baseline by 4.1 points in accuracy, 10.0 points in macro F1, 21.1\% in latency, and 22.7\% in energy. Source code is available at \href{https://anonymous.4open.science/r/Akida-Research-Project-F3AC}{Akida-Research-Project}.
\vspace{-4mm}
\end{abstract}
\vspace{-4mm}

\vspace{-8mm}
\section{Introduction}
\vspace{-4mm}
Skin lesions include benign conditions and malignant disease, and timely recognition is essential for early intervention and appropriate referral. In Australia, melanoma is among the most frequently diagnosed cancers, with 10,600 cases in males and 7,600 cases in females in 2023 \cite{aihw2024brief}. Cancer Council Australia further notes that Australia has one of the highest melanoma rates worldwide \cite{cancercouncil2025melanoma}.  This high burden increases the number of lesions requiring assessment and amplifies the clinical cost of delayed detection and incorrect triage. Because melanoma can closely resemble benign lesions in dermatoscopic appearance, there is a practical need for portable decision-support systems that perform multi-class triage at the point of care and help distinguish melanoma from visually similar benign conditions. Although deep learning-based diagnostic systems have achieved strong performance \cite{bhatt2023state}, many deployments remain cloud-centric and require transferring sensitive patient data \cite{yaqoob2023symmetry}, which increases security risk \cite{kourou2021applied} and is constrained by privacy regulation, including HIPAA \cite{mcgraw2021privacy} and GDPR \cite{kretschmer2021cookie}. For home and remote settings where connectivity and infrastructure are limited \cite{janda2022early}, practical systems must support on-device inference and efficient model adaptation without repeated cloud retraining.

\begin{figure}[t]
    \centering
    \includegraphics[width=0.99\linewidth]{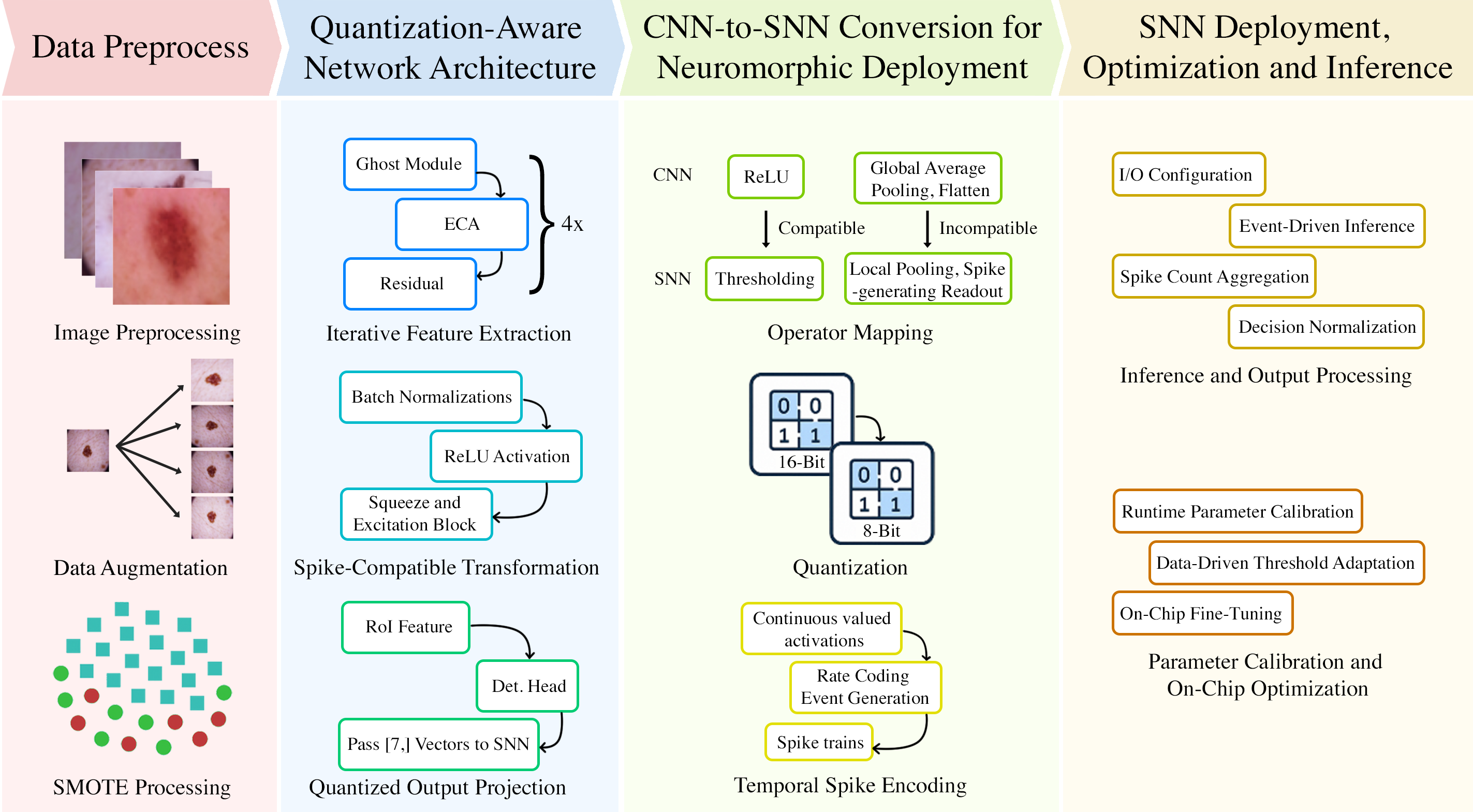}
    \vspace{-3mm}
    \caption{End-to-end framework for quantization-aware neuromorphic skin lesion classification: (1) lightweight image preprocessing using deterministic resize, padding, normalization, training-only augmentation, and embedding-space SMOTE; (2) a quantization-aware network for feature extraction and spike-compatible transformation; (3) CNN-to-SNN conversion with operator mapping and temporal spike encoding; and (4) SNN deployment with on-chip optimization for real-time and energy-efficient inference on edge hardware.}
    \vspace{-8mm}
    \label{fig:pipeline}
\end{figure}

Recently, Spiking Neural Networks (SNNs) and neuromorphic hardware have been explored as low-power alternatives for edge medical inference. SNNs encode information as sparse spike events, reducing unnecessary dense computation when activation activity is low \cite{han2022snnreview,kim2024snnbiomed}. Neuromorphic processors such as BrainChip Akida, IBM TrueNorth, and Intel Loihi provide event-driven execution models for resource-constrained deployment \cite{brainchip2025akida1000,akopyan2015truenorth,davies2018loihi}. A common route to neuromorphic deployment is to train a conventional CNN backbone such as ResNet or DenseNet and convert it to an SNN for edge inference \cite{he2016resnet,huang2017densenet,rueckauer2017conversion,hu2023fastsnn}. In practice, conversion is often brittle because standard CNN components such as BatchNorm, global pooling, and unbounded activations require careful folding, quantization, or operator lowering before they can be executed as integer spiking graphs \cite{jacob2018quantization,jiang2023conversion,hu2023fastsnn}. Accuracy can further degrade on small, imbalanced medical datasets because quantization error and reduced activation resolution may suppress rare lesion cues \cite{dablain2024imbalance}.

We propose a Quantization-Aware Neuromorphic Architecture (QANA) within the end-to-end pipeline in Fig.~\ref{fig:pipeline} to address these challenges. QANA consists of three modules. First, an iterative feature extraction and downsampling stage uses stacked Ghost blocks, spatially-aware efficient channel attention, and residual connections to reduce compute while preserving discriminative features required for dermatology. Second, a spike-compatible transformation stage applies quantization-aware normalization, bounded activations, and squeeze-and-excitation, where global reduction and channel gating are explicitly represented as quantized integer operations during conversion. Third, a quantized projection head maps the compact representation to class logits in a form that remains compatible with CNN-to-SNN conversion, low-power neuromorphic execution, and validation-based readout calibration.

On HAM10000, QANA achieves 91.6\% Top-1 accuracy and 91.0\% macro F1, improving the best converted SNN baseline by 3.5 points in Top-1 accuracy and 12.1 points in macro F1. On the clinical dataset, QANA attains 90.8\% Top-1 accuracy and 81.7\% macro F1, improving over the strongest baseline by 3.2 points in Top-1 accuracy and 3.6 points in macro F1. On BrainChip Akida, QANA processes one image in 1.5\,ms with 1.7\,mJ, outperforming the strongest size-matched Akida baseline by 4.1 points in Top-1 accuracy, 10.0 points in macro F1, 21.1\% in latency, and 22.7\% in energy.

The primary contributions of this research are as follows:
\vspace{-2mm}
\begin{itemize}
    \item We design QANA, a conversion-oriented quantization-aware architecture for dermatoscopic image classification that replaces conversion-fragile components with spike-compatible, bounded-activation blocks and lightweight attention to preserve discriminative features under 8-bit constraints.
    \item We formulate a graph-preserving CNN-to-SNN conversion pipeline for Akida that integrates BatchNorm reparameterization, uniform affine integer quantization, and deterministic operator lowering to ensure an all-integer executable spiking graph. Source code is available at \href{https://github.com/HaitianWang/Akida-Research-Project}{Akida-Research-Project}.
    \item We develop a deployment procedure on neuromorphic hardware that calibrates spike thresholds, integration windows, and the readout layer using validation data, enabling lightweight post-conversion readout calibration without full model retraining.
\end{itemize}

\vspace{-6mm}
\section{Related Work}
\vspace{-2mm}

Prior studies on neuromorphic hardware and SNNs have highlighted their suitability for edge inference in constrained environments \cite{han2022snnreview,kim2024snnbiomed}. Platforms such as BrainChip Akida, IBM TrueNorth, and Intel Loihi demonstrate event-driven execution models that reduce dense arithmetic and memory movement for low-power inference \cite{brainchip2025akida1000,akopyan2015truenorth,davies2018loihi}. These properties are attractive for medical image analysis on edge devices, where privacy, latency, and energy constraints are often coupled. However, practical dermatology deployment remains difficult because dermatoscopic datasets are often imbalanced, visually fine-grained, and sensitive to acquisition variability \cite{wu2022skin,bhatt2023state,dablain2024imbalance}.

CNN-to-SNN conversion has become a common strategy for transferring mature CNN architectures to neuromorphic processors \cite{rueckauer2017conversion,jiang2023conversion,hu2023fastsnn}. Conversion toolkits, including BrainChip CNN2SNN, support the mapping of quantized CNN graphs to Akida-compatible SNN models through supported operator lowering and runtime compilation \cite{brainchip2025cnn2snn,brainchip2025metatf}. However, conversion stability depends strongly on the operator set and activation statistics of the source CNN. Operations such as batch normalization, global reductions, nonlinear gates, and multi-bit activations require explicit folding, integer lowering, or scale management in integer-only conversion pipelines. If handled naively, these operations can introduce calibration error and amplify performance loss on imbalanced medical data \cite{jacob2018quantization,jiang2023conversion,dablain2024imbalance}.

In parallel, conventional CNN-based methods based on architectures such as DenseNet and Inception have shown strong performance on visual recognition and dermatological image analysis benchmarks \cite{huang2017densenet,szegedy2017inception,wu2022skin,bhatt2023state}. Despite their effectiveness, these models often require repeated retraining or recalibration to address domain shifts, new lesion types, or demographic variations \cite{wu2022skin,bhatt2023state,moor2023foundation}. Such retraining incurs latency, high computational cost, and data privacy concerns \cite{moor2023foundation}. Moreover, general CNN models often fail to generalize to underrepresented conditions such as rare tumors or images from diverse acquisition domains, highlighting the need for adaptive and privacy-preserving on-device solutions \cite{wu2022skin,bhatt2023state,dablain2024imbalance}. These limitations motivate the development of neuromorphic frameworks that maintain diagnostic performance while supporting incremental learning and efficient operation in resource-constrained settings.

\vspace{-3mm}
\section{Methodology}
\vspace{-3mm}

This section details the complete pipeline for the development and deployment of the proposed neuromorphic skin lesion classification system. As shown in Fig.~\ref{fig:pipeline}, the pipeline encompasses data preprocessing, the design of a quantization-aware neural network architecture compatible with spiking inference, conversion to an event-driven SNN format, and hardware-level deployment on the Akida platform.

\vspace{-6mm}
\subsection{Data Preprocessing}
\vspace{-2mm}

The pipeline includes three main stages: image preprocessing to ensure quality and compatibility with the Akida hardware, data augmentation to enhance dataset diversity and robustness, and SMOTE to address severe class imbalance among lesion categories.

\vspace{-2mm}
\subsubsection{Image Preprocessing} 
\vspace{-2mm} 

All dermatoscopic images undergo automated integrity checks followed by a brief manual review. Inputs are processed at $64\times64$ to match the Akida on-chip memory budget used in our deployment. To avoid adding a separate lesion localization network before neuromorphic inference, preprocessing is restricted to deterministic image operations. Each image is converted to RGB, padded to a square canvas while preserving its aspect ratio, and resized to $64\times64$ using antialiased area resampling. Pixel intensities are then normalized per channel to $[0,1]$. No saliency model, external detector, segmentation network, denoising, inpainting, or class-dependent cropping is used. The same preprocessing rule is applied to training, validation, and test images, ensuring that the input pipeline is independent of labels, split statistics, and test-time outcomes.

\vspace{-5mm}
\subsubsection{Data Augmentation}
\vspace{-1mm}
Augmentations preserve dermoscopic color cues. We use random horizontal and vertical flips (probability 0.5). Luminance is perturbed on the CIELAB $L$ channel with a multiplicative factor in $[0.95,1.05]$ or a gamma factor in $[0.97,1.03]$. Contrast is adjusted in $[0.95,1.05]$. No hue shift is applied. A gray-world normalization fitted on the training split is used during training only. All random draws use fixed seeds for reproducibility.

\begin{figure}[t]
    \centering
    \includegraphics[width=1\linewidth]{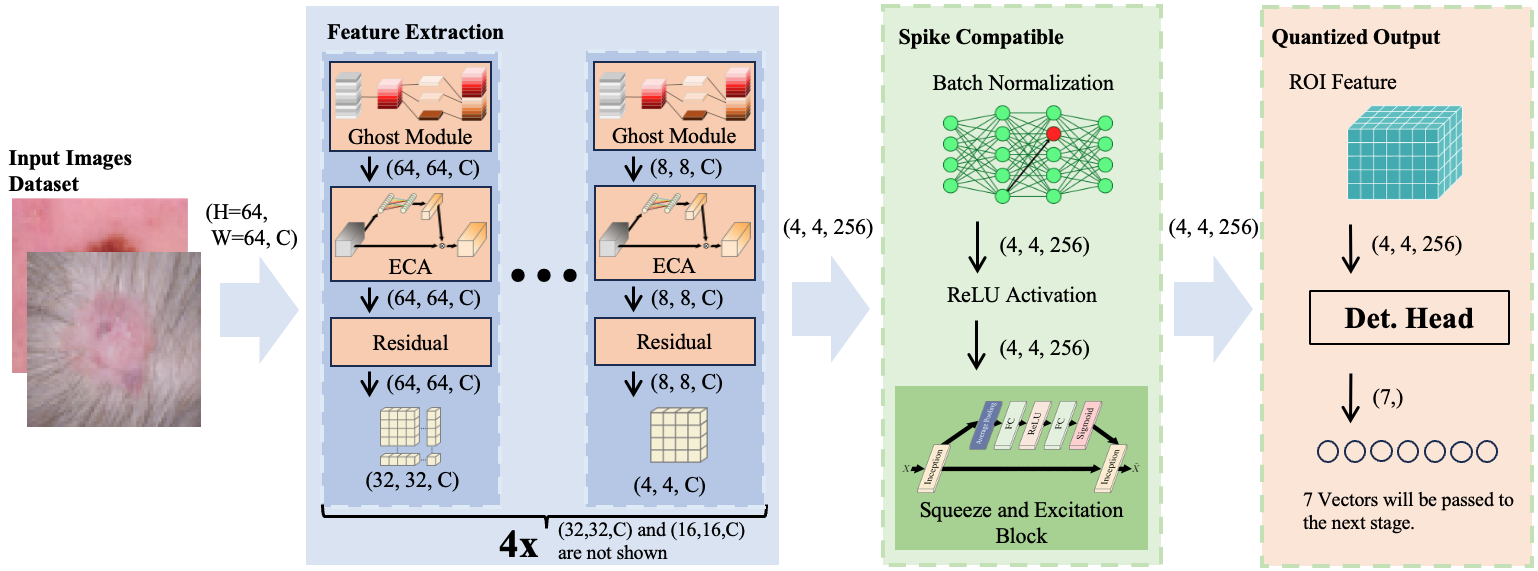}
    \vspace{-9mm}
    \caption{Architecture of QANA. The backbone performs iterative feature extraction using stacked Ghost modules \cite{han2020ghostnet}, spatially-aware efficient channel attention \cite{wang2020eca}, and residual blocks. A spike-compatible transformation stage applies bounded activation and squeeze-and-excitation \cite{hu2018senet} before quantized projection to class logits.}
    \label{fig:network}
    \vspace{-7mm}
\end{figure}

\vspace{-3mm}
\subsubsection{Synthetic Minority Oversampling (SMOTE)}
\vspace{-1mm}
Following SMOTE \cite{chawla2002smote}, we perform minority oversampling strictly in the learned embedding space rather than on raw pixels. For each training image $x_i$ from class $c$, the backbone and spike-compatible transform produce a penultimate representation $\mathbf{r}_i \in \mathbb{R}^{4096}$ after global flattening. Let $\mathcal{I}_c$ be the index set of class $c$ and $\{\mathbf{r}_i\}_{i\in\mathcal{I}_c}$ the corresponding embeddings. Before neighbor search, embeddings are standardized on the \emph{training split only} to avoid information leakage. A PCA whitening is then fitted on the same split and applied to stabilize distances in high dimension. Writing the sample covariance as $\mathbf{\Sigma}=\mathbf{U}\mathbf{\Lambda}\mathbf{U}^\top$ and retaining the top $d=256$ components, the whitened vector is
\vspace{-2mm}
\begin{small}
\vspace{-1mm}
\begin{equation}
\widetilde{\mathbf{r}}_i=\mathbf{\Lambda}_d^{-\frac{1}{2}}\mathbf{U}_d^\top(\mathbf{r}_i-\boldsymbol{\mu}),
\label{eq:whitening}
\end{equation}
\end{small}
where $\boldsymbol{\mu}$ is the training mean, $\mathbf{U}_d\in\mathbb{R}^{4096\times d}$ the leading eigenvectors and $\mathbf{\Lambda}_d\in\mathbb{R}^{d\times d}$ their eigenvalues. Nearest neighbors are computed with Euclidean distance in whitened space; for each $i\in\mathcal{I}_c$ we form a $k$-NN set $\mathcal{N}_i$ with $k=5$.


To reduce the influence of outliers, we keep only ``safe'' minority anchors whose local median neighbor distance is below the class-specific 90th percentile:
\begin{equation}
\begin{aligned}
\mathcal{S}_c^{\text{safe}}
=\Big\{\, i\in\mathcal{I}_c \ \Big|\
&\operatorname{med}_{j\in\mathcal{N}_i}\big\lVert\widetilde{\mathbf r}_i-\widetilde{\mathbf r}_j\big\rVert_2
&\le
Q_{0.9}\!\left(
\big\{\operatorname{med}_{j\in\mathcal{N}_u}\big\lVert\widetilde{\mathbf r}_u-\widetilde{\mathbf r}_j\big\rVert_2\big\}_{u\in\mathcal{I}_c}
\right)
\Big\}.
\end{aligned}
\label{eq:safe_set}
\end{equation}

For each minority class $c$ with $N_c=|\mathcal{I}_c|$ and $N_{\max}=\max_{c'}N_{c'}$, we synthesize to a target level $\tau N_{\max}$ with $\tau=0.85$ to avoid distribution collapse. The number of synthetic embeddings is
\begin{small}
\begin{equation}
M_c=\max\!\big(0,\;\lfloor \tau N_{\max}-N_c \rfloor\big).
\label{eq:quota}
\end{equation}
\end{small}
Each synthetic vector is generated from an anchor $i\in\mathcal{S}_c^{\text{safe}}$ and a random neighbor $j\in\mathcal{N}_i$ by a convex interpolation and a small, locally aligned perturbation to improve diversity. Let $\mathbf{C}_i$ be the empirical covariance of $\{\widetilde{\mathbf{r}}_u\}_{u\in\{i\}\cup\mathcal{N}_i}$, and let $\mathbf{U}_i^{(m)}$ contain its $m=8$ principal directions. With $\delta_i=\operatorname{med}_{j\in\mathcal{N}_i}\|\widetilde{\mathbf{r}}_i-\widetilde{\mathbf{r}}_j\|_2$ and $\sigma_c=0.05\,\operatorname{med}_{u\in\mathcal{I}_c}\delta_u$, we sample
\begin{small}
\begin{equation}
\widetilde{\mathbf{r}}_{\text{new}}
=\widetilde{\mathbf{r}}_i+\eta\big(\widetilde{\mathbf{r}}_{j}-\widetilde{\mathbf{r}}_i\big)
+\sigma_c\,\mathbf{U}_i^{(m)}\boldsymbol{\xi},\quad
\eta\sim\mathcal{U}(0,1),\;\boldsymbol{\xi}\sim\mathcal{N}(\mathbf{0},\mathbf{I}_m).
\label{eq:gen_rule}
\end{equation}
\end{small}
The synthetic embedding is then mapped back to the original feature space by the inverse whitening:
\begin{small}
\begin{equation}
\mathbf{r}_{\text{new}}=\boldsymbol{\mu}+\mathbf{U}_d\mathbf{\Lambda}_d^{\frac{1}{2}}\widetilde{\mathbf{r}}_{\text{new}}.
\label{eq:inverse_whiten}
\end{equation}
\end{small}

Synthetic data are \emph{not} images. They are injected at the penultimate layer as additional feature vectors and used to train the quantized output projection with cross-entropy. During this stage the backbone up to the spike-compatible transform is frozen; only the classifier head is updated for 5 epochs with mixed real and synthetic embeddings. A subsequent short fine-tuning on real data re-enables the last block and the head to reconcile the feature distribution. Input images are still resized to $64\times64$ prior to feature extraction for hardware constraints, but no pixel-space mixing or image synthesis is performed at any time.

\begin{figure}[t]
    \centering
    \vspace{-4mm}
    \includegraphics[width=0.75\linewidth]{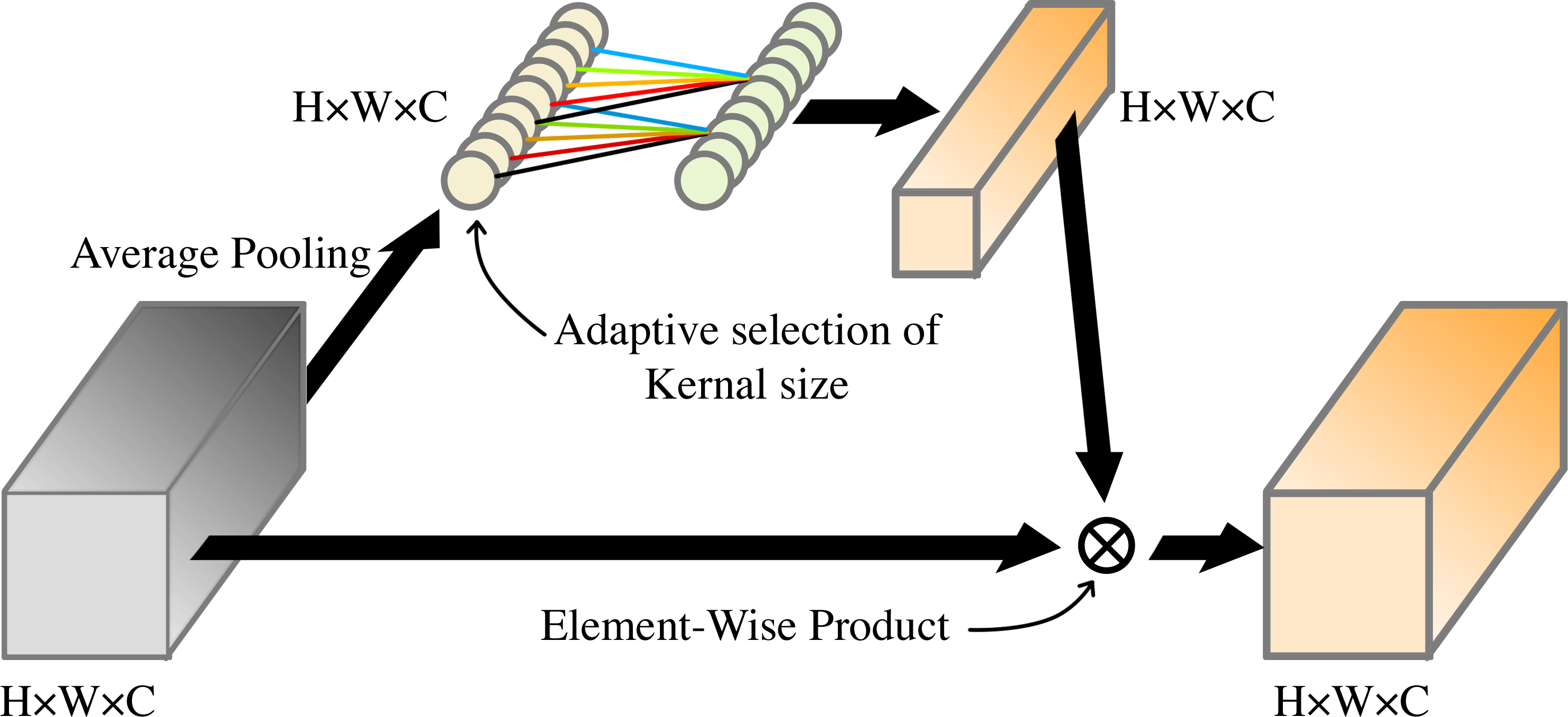}
    \vspace{-2mm}
    \caption{Spatially-Aware ECA variant used in QANA. The input feature map is processed by depthwise spatial aggregation, batch normalization, and a $1\times1$ channel-wise projection. A sigmoid gate produces channel weights, which are applied through channel-wise multiplication to generate the output feature map.}
    \vspace{-6mm}
    \label{fig:eca-block}
\end{figure}

\vspace{-6mm}
\subsection{Quantization-Aware Network Architecture}
\label{sec:hierarchical_backbone}
\vspace{-2mm}

As shown in Fig.~\ref{fig:network}, QANA follows a hierarchical design with four lightweight feature extraction stages and a spike-compatible transformation stage that produces SNN-ready activations for deterministic CNN-to-SNN conversion on Akida. The backbone prioritizes conversion stability by constraining activation ranges and by using operators that admit integer lowering, while preserving discriminative capacity through multi-scale feature generation and channel recalibration.

\noindent\textbf{Backbone configuration.}
The input is $64\times64\times3$. The backbone contains four stages, each consisting of a Ghost block followed by BN, ReLU6, and a residual shortcut. Spatial downsampling is performed by $2\times2$ max pooling with stride $2$ at the end of each stage, yielding a $4\times4$ feature map after stage four. When the shortcut changes the channel dimension, a $1\times1$ projection is used to match dimensions before residual addition. Dropout is applied only during training after the bounded activation to reduce overfitting on minority classes.

\vspace{-4mm}
\subsubsection{Iterative Feature Extraction and Downsampling}
\vspace{-1mm}

Feature extraction is implemented as a stack of Ghost blocks \cite{han2020ghostnet} that expand representational diversity under a tight FLOP budget. For stage $l\in\{1,\ldots,4\}$, the input tensor $F^{(l-1)}$ is first mapped to a compact set of primary features using a pointwise projection. Additional features are then generated using depthwise transformations, and the two parts are concatenated to form the output feature map:

\begin{small}
\vspace{-1mm}
\begin{equation}
\begin{aligned}
F_{\mathrm{p}}^{(l)} &= \mathcal{F}_{1\times 1}^{(l)}\!\left(F^{(l-1)}\right),
F_{\mathrm{g}}^{(l)} &= \mathcal{F}_{\mathrm{dw}}^{(l)}\!\left(F_{\mathrm{p}}^{(l)}\right),
F_{\mathrm{ghost}}^{(l)} &= \mathrm{Concat}\!\left(F_{\mathrm{p}}^{(l)},\,F_{\mathrm{g}}^{(l)}\right).
\end{aligned}
\label{eq:ghost_complex}
\end{equation}
\end{small}
In practice, each stage uses a fixed Ghost expansion ratio so that the primary projection produces a compact subset of channels and the remaining channels are generated by depthwise operations. This allocation concentrates most computation in pointwise and depthwise convolutions while preserving channel diversity under a small FLOP budget. After feature generation, each stage applies BatchNorm followed by a bounded ReLU6 activation to keep post-activation values within a fixed and quantization-friendly range, then applies dropout for regularization. This bounded design reduces calibration sensitivity during 8-bit quantization and improves conversion robustness when activations are later encoded into spikes.

To model channel dependencies while avoiding repeated global reductions in the feature extraction stages, each stage integrates a Spatially-Aware Efficient Channel Attention (SA-ECA) variant inspired by ECA \cite{wang2020eca}. As illustrated in Fig.~\ref{fig:eca-block}, SA-ECA uses depthwise spatial aggregation, batch normalization, and a $1\times1$ channel-wise projection to produce per-channel weights, followed by a sigmoid gate and channel-wise multiplication. This design keeps the attention path local and compact, making its convolutional and scaling operations suitable for integer lowering during CNN-to-SNN conversion.

To maintain stable deep stacking in small-data regimes, each stage uses a residual shortcut with optional projection for dimension alignment, followed by spatial downsampling via max pooling. After four stages, the spatial resolution is reduced to $4\times4$, producing a compact feature tensor suitable for spike-compatible transformation.


\vspace{-4mm}
\subsubsection{Spike-Compatible Feature Transformation}
\vspace{-1mm}

The backbone output $F^{(4)}$ is passed to a spike-compatible transformation module designed for direct quantization and SNN integration. We first apply a SeparableConv2D $(3\times3,256)$, following the depthwise-separable convolution design used in efficient CNNs \cite{chollet2017xception,sandler2018mobilenetv2}, to increase channel capacity while preserving hardware efficiency, then apply BatchNorm and an activation constraint that clips values into $[0,1]$:
\begin{small}
\vspace{-2mm}
\begin{align}
    \widehat{F} = \min\left(1, \max\left(0, \gamma_{\text{spk}} \cdot \text{BN}(\text{SepConv}_{3\times 3, 256}(F^{(4)})) + \beta_{\text{spk}}\right)\right)
    \label{eq:spike_quant_complex}
\end{align}
\end{small}
This bounded mapping aligns the activation range with spike encoding and avoids saturation under 8-bit affine quantization used in the subsequent conversion pipeline.

To further recalibrate channel responses, QANA applies a Squeeze-and-Excitation (SE) block \cite{hu2018senet} after the bounded spike-compatible activation. The SE module computes channel descriptors by global average pooling, passes them through a two-layer bottleneck with a reduction ratio of $16$, and rescales $\widehat{F}$ using quantized sigmoid gates. Although SE contains a global reduction, it is not treated as an unsupported floating-point operator. In the conversion graph, global average pooling is lowered to an integer spatial reduction, the two fully connected layers are emitted as quantized integer linear projections, and the final channel-wise rescale is folded into the following pointwise projection when possible. Otherwise, it remains an integer per-channel scaling operation. This design keeps channel recalibration compatible with the all-integer graph while improving sensitivity to minority lesion classes.

\vspace{-4mm}
\subsubsection{Quantized Output Projection}
\vspace{-2mm}

Finally, the spike-compatible tensor is flattened to a vector $r\in\mathbb{R}^{4096}$ and projected to class logits through a linear readout:
\vspace{-2mm}
\begin{small}
\vspace{-2mm}
\begin{equation}
\mathbf{y} = \mathbf{W}_{\text{cls}}\, r + \mathbf{b}_{\text{cls}},
\label{eq:detection_head}
\end{equation}
\end{small}
where $\mathbf{W}_{\text{cls}}\in\mathbb{R}^{7\times 4096}$ and $\mathbf{b}_{\text{cls}}\in\mathbb{R}^{7}$. This projection is quantization-aware during training and produces logits in a form that is directly consumable by the CNN-to-SNN converter for neuromorphic inference.

\vspace{-6mm}
\subsection{CNN-to-SNN Conversion for Neuromorphic Deployment}
\label{sec:cnn2snn}
\vspace{-2mm}

We convert the trained quantization-aware CNN to an event-driven SNN with a graph-preserving flow that enforces an all-integer executable graph on Akida. The conversion consists of (i) static reparameterization, (ii) 8-bit affine quantization, and (iii) deterministic operator lowering.

\textbf{BN folding and activation constraint.}
All BatchNorm layers are folded into the preceding convolution or pointwise projection. For output channel $c$ with convolution weights $\mathbf{W}_c$ and bias $b_c$ and BN parameters $(\gamma_c,\beta_c,\mu_c,\sigma_c^2,\epsilon)$, we use
\begin{small}
\begin{equation}
\mathbf{W}_c'=\frac{\gamma_c}{\sqrt{\sigma_c^2+\epsilon}}\mathbf{W}_c,\qquad
b_c'=\frac{\gamma_c}{\sqrt{\sigma_c^2+\epsilon}}\left(b_c-\mu_c\right)+\beta_c .
\label{eq:bn_fuse}
\end{equation}
\end{small}
Depthwise convolutions are folded channel-wise in the same manner. During training, non-supported nonlinearities are replaced by bounded ReLU6 so that post-activation ranges are fixed and compatible with spike encoding \cite{jacob2018quantization,rueckauer2017conversion}.

\vspace{-0mm}
\textbf{8-bit affine quantization.}
Weights and activations are quantized with uniform affine 8-bit fixed-point. For a real activation $a$ and calibration-derived scale and zero-point $(s_a,z_a)$, the quantized value is
\vspace{-2mm}
\begin{small}
\vspace{-2mm}
\begin{equation}
a_q=\mathrm{clip}\!\left(\mathrm{round}\!\left(\frac{a}{s_a}\right)+z_a,\,0,\,255\right),\qquad
a\approx \left(a_q-z_a\right)s_a .
\label{eq:affine_quant}
\end{equation}
\end{small}
For ReLU6-bounded activations, we set $z_a=0$ and $s_a=6/255$, which maps $[0,6]$ to $[0,255]$ without saturation. For layers constrained to $[0,1]$, including Eq.~\ref{eq:spike_quant_complex}, we set $z_a=0$ and $s_a=1/255$.

\textbf{Spike emission.}
SNN inference uses an integrate-and-fire neuron with subtractive reset driven by integer activations. With membrane state $u_t$, integer drive $a_q$, and threshold $\tau$, spike generation follows
\vspace{-1mm}
\begin{small}
\vspace{-1mm}
\begin{equation}
s_t=\mathbb{I}\!\left(u_{t-1}+a_q\ge \tau\right),\qquad
u_t=\left(u_{t-1}+a_q\right)-\tau s_t .
\label{eq:spike_emit}
\end{equation}
\end{small}
Class scores are obtained from spike counts over the integration window, and Softmax is applied after spike integration \cite{rueckauer2017conversion}.

\textbf{Integer graph lowering and metrics.}
Supported operators are lowered deterministically to integer kernels, including convolution, depthwise convolution, separable convolution, pooling, elementwise clipping, residual addition with scale alignment, channel-wise rescaling, and linear layers. For residual addition, the two integer tensors are brought to a common affine scale before summation. Given $(s_1,z_1)$ and $(s_2,z_2)$, we choose a target scale $s_\oplus=\max(s_1,s_2)$ and rescale each branch by an integer multiplier and right shift so that both branches represent values under $s_\oplus$ prior to addition. This prevents overflow and reduces drift introduced by mixed-scale accumulation. Global average pooling in the SE block is compiled as an integer spatial reduction with the division absorbed into the output scale. The two SE fully connected layers are emitted as quantized integer linear layers, and the sigmoid gate is implemented through a quantized lookup or an equivalent bounded gate approximation. The resulting channel-wise rescale is folded into the next pointwise projection when possible, otherwise it remains an integer per-channel scaling operation. Reshape operations do not introduce compute. For the final QANA model, any subgraph requiring floating-point dequantization is rejected during strict deployment. For diagnostic ablations, we report the proportion of operators that satisfy strict integer lowering before compatibility repair, which is used only to quantify conversion robustness. We report conversion success as
$r_{\mathrm{succ}} = 1 - \lvert \mathcal{V}_{\mathrm{fp32}} \rvert / \lvert \mathcal{V} \rvert$,
and quantify conversion-induced changes by $\Delta\mathrm{Acc}=\mathrm{Acc}_{\mathrm{SNN}}-\mathrm{Acc}_{\mathrm{CNN}}$ and $\Delta\mathrm{F1}=\mathrm{F1}_{\mathrm{SNN}}-\mathrm{F1}_{\mathrm{CNN}}$ on identical splits and preprocessing.

\vspace{-3mm}
\subsection{SNN Deployment, Optimization and Inference}
\vspace{-2mm}

The converted SNN model is deployed directly onto the BrainChip Akida neuromorphic processor for hardware-based inference. The deployment process comprises several steps: loading the SNN model into the hardware runtime environment, configuring input/output data streams, and initializing internal buffers and neuron state registers.

\vspace{-4mm}
\subsubsection{Inference and Output Processing}
\vspace{-1mm}

Inference is performed in an event-driven manner, with input images encoded into spike trains and propagated through the SNN in a fully parallel fashion. For each test or validation sample, the model outputs spike counts or firing rates at the final output neurons, corresponding to the target classes. To robustly map temporal spike responses to class probabilities, we aggregate the spike counts $S_c(t)$ for each class $c$ within an integration window $T$, and apply a soft decision normalization:
\begin{small}
\begin{align}
    \hat{p}_c = \frac{\exp\left(\alpha\, \sum_{t=1}^{T} w_t S_c(t)\right)}{\sum_{k=1}^{C} \exp\left(\alpha\, \sum_{t=1}^{T} w_t S_k(t)\right)}
    \label{eq:softmax_spike_decision}
\end{align}
\end{small}
where $w_t$ is an optional temporal weighting factor (e.g., $w_t = \exp(-\beta (T-t))$ for decaying integration), $\alpha$ is a scaling parameter, and $C$ is the number of output classes. Class prediction is then made as $\arg\max_c \hat{p}_c$. 

In our implementation, we use uniform temporal weights $w_t=1$ unless validation indicates a benefit from decaying integration. The scaling factor $\alpha$ is selected on the validation split and then fixed for the reported test evaluation.

\vspace{-4mm}
\subsubsection{Parameter Calibration and On-Chip Optimization}

After initial deployment, key runtime parameters, including output spike thresholds, integration windows, and temporal pooling parameters, are empirically calibrated on a held-out validation set. To further optimize class assignment and suppress spurious output events, threshold adaptation is performed per class: 
\begin{align}
    \theta_c^{*} =
    \arg\min_{\theta}
    \sum_{i=1}^{N}
    \mathbb{I}\left[
    \mathbb{I}(y_i=c) \neq \mathbb{I}\left(S_c^{(i)}>\theta\right)
    \right].
    \label{eq:adaptive_threshold}
\end{align}
where $S_c^{(i)}$ is the total spike count for class $c$ on sample $i$, $y_i$ is the true label, and $\mathbb{I}[\cdot]$ is the indicator function. This calibration changes only the output decision thresholds and does not introduce additional image crops or auxiliary model inference. If performance deviation is observed relative to the original CNN, lightweight readout calibration is performed by adjusting only the last output layer while keeping the feature extractor fixed. All calibration, threshold selection, and optional on-chip fine-tuning steps use only the validation split. The test split is used only for final reporting and is not involved in calibration, threshold selection, or readout adjustment.


\vspace{-6mm}
\section{Experiment}
\vspace{-4mm}

This section presents the datasets, experimental setup, and evaluation protocols, followed by detailed quantitative analyses of classification performance, ablation studies, and efficiency metrics under various deployment conditions.

\vspace{-6mm}
\subsection{Datasets}
\vspace{-2mm}

We used two datasets: the public HAM10000 benchmark and a proprietary clinical dataset from Nanjing Drum Tower Hospital, China. Both sets reflect a range of lesion types and clinical diversity.

\vspace{-4mm}
\subsubsection{HAM10000 Dataset}
\vspace{-2mm}

The HAM10000 dataset~\cite{tschandl2018ham10000} consists of 10,015 dermatoscopic RGB images labeled by expert dermatologists into 7 categories: melanocytic nevus, melanoma, benign keratosis-like lesions, basal cell carcinoma, actinic keratoses and intraepithelial carcinoma, vascular lesions, and dermatofibroma. The images originate from diverse sources and exhibit significant class imbalance. For all experiments, we adopted a standard 70\%/10\%/20\% train/validation/test split.

\vspace{-4mm}
\subsubsection{Hospital Clinical Dataset}
\vspace{-2mm}
A proprietary clinical dataset was established through a research collaboration with Nanjing Drum Tower Hospital, comprising 3,162 dermatoscopic images from 1,235 patients collected between June 2022 and February 2024 (Project No. NJG-2022-3233-CN). To ensure comparability with benchmark studies, lesion categories were curated to match the seven classes in HAM10000. Rare and unclassified conditions were excluded to maintain consistent labeling across classes. Each case was independently annotated by at least 2 board-certified dermatologists, with histopathological confirmation available for a subset of cases. All images and associated metadata were de-identified in accordance with ethics and governance requirements. Due to patient privacy, the dataset is not publicly released. Dataset partitioning followed a 70\%/10\%/20\% train/validation/test split stratified by disease category. Splits were constructed to avoid patient overlap across training, validation, and test sets.

\begin{table}[t]
\vspace{-4mm}
\caption{Class-wise precision, recall, F1 score of QANA on the HAM10000 test set.}
\vspace{-3mm}
\label{tab:ours_performance}
\centering
\setlength{\tabcolsep}{5pt}
\begin{tabular}{lccc}
\toprule[1.5pt]  
\textbf{Class} & \textbf{Precision} & \textbf{Recall} & \textbf{F1} \\
\hline
Actinic keratoses  & 0.890 & 0.933 & 0.911 \\
Basal cell carcinoma                               & 0.890 & 0.901 & 0.896 \\
Benign keratosis-like lesions                        & 0.866 & 0.853 & 0.859 \\
Dermatofibroma                                      & 0.925 & 0.976 & 0.950 \\
Melanocytic nevi                                     & 0.887 & 0.817 & 0.851 \\
Vascular lesions                                  & 0.949 & 0.966 & 0.957 \\
Melanoma                                           & 0.956 & 0.933 & 0.944 \\
\hline
\textbf{Macro average}                                              & \textbf{0.909} & \textbf{0.911} & \textbf{0.910} \\
\bottomrule[1.5pt]
\end{tabular}
\vspace{-2mm}
\end{table}

\begin{table}[t]
\vspace{-1mm}
\caption{Performance Comparison of Converted SNN Models on HAM10000}
\vspace{-3mm}
\label{tab:ham10000_comparison}
\centering
\renewcommand{\arraystretch}{1.12}
\begin{tabular}{lcc}
\toprule[1.5pt] 
\textbf{Model (SNN, Akida)}           & \textbf{Top-1 Accuracy (\%)} & \textbf{Macro F1 (\%)} \\
\hline
ResNet-50~\cite{he2016resnet}                             & 85.7                         & 76.4 \\
DenseNet-121~\cite{huang2017densenet}                     & 86.5                         & 77.2 \\
Inception-v4~\cite{szegedy2017inception}                  & 85.9                         & 76.9 \\
EfficientNet-B4~\cite{tan2019efficientnet}                & 87.3                         & 78.1 \\
MobileNet-v2~\cite{sandler2018mobilenetv2}                & 83.4                         & 74.7 \\
SENet-154~\cite{hu2018senet}                              & 86.9                         & 77.8 \\
Xception~\cite{chollet2017xception}                       & 85.5                         & 76.2 \\
Multi-Scale Attention~\cite{qian2022gmab}                 & 87.0                         & 78.0 \\
CNN Ensemble~\cite{harangi2018ensemble}                   & 88.1                         & 78.9 \\
AKIDANet~\cite{brainchip2025akidanet}                     & 83.2                         & 73.6 \\
\textbf{Ours}            & \textbf{91.6}                & \textbf{91.0} \\
\hline
\end{tabular}
\vspace{-7mm}
\end{table}

\vspace{-6mm}
\subsection{Experimental Setup}
\vspace{-2mm}

All model training, validation, and CNN-to-SNN conversion were conducted on a workstation with an Intel Core i9-12900K CPU, 128\,GB RAM, and NVIDIA RTX 3090 GPU, running Ubuntu 22.04 LTS. The software environment included Python 3.9, CUDA 11.8, TensorFlow 2.10, and Akida MetaTF SDK v2.2.1, including QuantizeML, CNN2SNN, and the Akida runtime \cite{brainchip2025metatf,brainchip2025cnn2snn}. Neuromorphic inference was performed on a BrainChip Akida AKD1000 PCIe board installed in the same system, with deployment and testing managed via the Akida Python API and default board settings. 


\vspace{-5mm}
\subsection{Analysis of Classification Results on HAM10000 Dataset}
\vspace{-2mm}

Table~\ref{tab:ours_performance} reports the class-wise precision, recall, and F1 score of QANA on the HAM10000 test set. Macro F1 is computed as the unweighted mean of per-class F1 scores on the test set. The model achieves consistent performance across lesion categories, including minority classes. With a Top-1 accuracy of 91.6\% and a macro F1 of 91.0\%, the system demonstrates effective discrimination of both common and rare lesions.

\begin{table}[t]
\vspace{-2mm}
\caption{
Per-image inference latency and energy consumption of all models on the HAM10000 test set. CNN baselines are measured on an NVIDIA RTX 3090 (GPU) and Intel i9-12900K (CPU); SNNs are measured on BrainChip Akida AKD1000. All values are averaged over 10{,}000 single-image inference runs sampled from the fixed test split after warm-up. The rightmost columns show the percentage reduction achieved by Akida SNN inference relative to the corresponding GPU-based CNN implementation.}
\vspace{-4mm}
\label{tab:latency_energy}
\centering
\small
\setlength{\tabcolsep}{3.2pt}
\renewcommand{\arraystretch}{1.08}
\resizebox{\textwidth}{!}{%
\begin{tabular}{lcccccccc}
\toprule[1.5pt]
\textbf{Model} & \multicolumn{2}{c}{\textbf{CNN (GPU)}} & \multicolumn{2}{c}{\textbf{CNN (CPU)}} & \multicolumn{2}{c}{\textbf{SNN (Akida)}} & \multicolumn{2}{c}{\textbf{Relative Reduction (\%)}} \\
\cline{2-9}
 & Latency (ms) & Energy (mJ) & Latency (ms) & Energy (mJ) & Latency (ms) & Energy (mJ) & Latency & Energy \\
\hline
ResNet-50~\cite{he2016resnet}           & 12.1  & 175.2 & 57.9  & 923.3  & 2.8 & 3.3 & 76.9 & 98.1 \\
DenseNet-121~\cite{huang2017densenet}         & 14.7  & 199.6 & 68.4  & 1075.2 & 3.1 & 3.5 & 78.9 & 98.2 \\
Inception-v4~\cite{szegedy2017inception}    & 16.8  & 218.5 & 82.1  & 1237.5 & 3.5 & 4.1 & 79.2 & 98.1 \\
EfficientNet-B4~\cite{tan2019efficientnet}        & 18.9  & 242.1 & 93.6  & 1345.6 & 4.0 & 4.6 & 78.8 & 98.1 \\
MobileNet-v2~\cite{sandler2018mobilenetv2} & \textbf{6.9} & \textbf{97.5} & 68.2 & 1082.1 & 2.2 & 2.6 & 68.1 & 97.3 \\
SENet-154~\cite{hu2018senet} & 17.3 & 236.8 & 89.4 & 1312.4 & 3.8 & 4.4 & 78.0 & 98.1 \\
Xception~\cite{chollet2017xception}         & 10.7 & 151.2 & \textbf{51.5} & \textbf{863.7} & 2.6 & 3.1 & 75.7 & 97.9 \\
Multi-Scale Attention~\cite{qian2022gmab}   & 19.8 & 251.7 & 97.7 & 1391.5 & 4.2 & 5.0 & 78.8 & 98.0 \\
CNN Ensemble~\cite{harangi2018ensemble}         & 36.1 & 450.5 & 157.2 & 2177.1 & 7.5 & 8.6 & 79.2 & 98.1 \\
\hline
\textbf{QANA (Ours)}                         & 27.6 & 163.1 & 83.9 & 841.5 & \textbf{1.5} & \textbf{1.7} & \textbf{94.6} & \textbf{99.0} \\
\bottomrule[1.5pt]
\end{tabular}}
\vspace{-3mm}
\end{table}

\begin{table}[t]
\caption{Ablation study of core modules in our model on the HAM10000 test set. Metrics are reported as percentages (\%). Each row shows the incremental addition of modules. Calib.\ denotes validation-based readout calibration after conversion, while evaluation is performed on the fixed test split.}
\vspace{-4mm}
\label{tab:ablation_study}
\centering
\small
\setlength{\tabcolsep}{2.6pt}
\renewcommand{\arraystretch}{1.08}
\resizebox{\textwidth}{!}{%
\begin{tabular}{lcccccc|ccccc}
\toprule[1.5pt]
\textbf{Configuration} & \textbf{Ghost} & \textbf{ECA} & \textbf{SE} & \textbf{Aug.} & \textbf{SMOTE} & \textbf{Calib.} & \textbf{Acc.} & \textbf{Rec.} & \textbf{Prec.} & \textbf{F1} & \textbf{AUC} \\
\midrule
Baseline                &  &  &  &  &  &  & 74.1 & 71.4 & 71.9 & 71.6 & 77.3 \\
+ Ghost Block           & \checkmark &  &  &  &  &  & 72.3 & 70.2 & 70.0 & 70.6 & 70.9 \\
+ ECA                   & \checkmark & \checkmark &  &  &  &  & 88.7 & 85.8 & 87.2 & 86.5 & 90.7 \\
+ SE                    & \checkmark & \checkmark & \checkmark &  &  &  & 89.8 & 87.7 & 88.1 & 87.8 & 91.5 \\
+ Augmentation          & \checkmark & \checkmark & \checkmark & \checkmark &  &  & 90.4 & 88.1 & 89.1 & 88.6 & 92.1 \\
+ SMOTE                 & \checkmark & \checkmark & \checkmark & \checkmark & \checkmark &  & 91.0 & 89.2 & 90.0 & 89.6 & 92.7 \\
+ Readout Calibration  & \checkmark & \checkmark & \checkmark & \checkmark & \checkmark & \checkmark & \textbf{91.6} & \textbf{91.1} & \textbf{90.9} & \textbf{91.0} & \textbf{93.4} \\
\bottomrule
\end{tabular}}
\vspace{-8mm}
\end{table}

\vspace{-5mm}
\subsection{Classification Performance on HAM10000}
\vspace{-2mm}

To evaluate neuromorphic skin lesion classification under consistent deployment constraints, we trained a set of widely used CNN backbones and converted each model to an SNN using the same quantization-aware pipeline, then executed all converted networks on the same Akida platform with identical preprocessing and runtime settings. Table~\ref{tab:ham10000_comparison} summarizes the resulting Akida-SNN performance across canonical and advanced baselines, which typically exhibit reduced accuracy and macro F1 after conversion, with Top-1 accuracy spanning 83.2\% to 88.1\%. QANA attains 91.6\% Top-1 accuracy and 91.0\% macro F1 on HAM10000, yielding the best overall results under the same neuromorphic execution setting.

\begin{table}[t]
\vspace{-3mm}
\caption{Performance comparison of converted SNN models on the Clinical Dataset.}
\vspace{-3mm}
\label{tab:your_dataset_comparison}
\centering
\renewcommand{\arraystretch}{1.12}
\begin{tabular}{lcc}
\toprule[1.5pt] 
\textbf{Model (SNN, Akida)}    & \textbf{Top-1 Accuracy (\%)} & \textbf{Macro F1 (\%)} \\
\hline
ResNet-50~\cite{he2016resnet}                & 84.6                         & 75.3 \\
DenseNet-121~\cite{huang2017densenet}        & 85.7                         & 76.2 \\
Inception-v4~\cite{szegedy2017inception}     & 85.2                         & 75.8 \\
EfficientNet-B4~\cite{tan2019efficientnet}   & 86.3                         & 77.0 \\
MobileNet-v2~\cite{sandler2018mobilenetv2}   & 82.8                         & 73.7 \\
SENet-154~\cite{hu2018senet}                 & 85.4                         & 76.6 \\
Xception~\cite{chollet2017xception}          & 84.2                         & 74.8 \\
Multi-Scale Attention~\cite{qian2022gmab}    & 86.5                         & 77.2 \\
CNN Ensemble~\cite{harangi2018ensemble}      & 87.6                         & 78.1 \\
AKIDANet~\cite{brainchip2025akidanet}        & 81.9                         & 71.5 \\
\textbf{Ours}                  & \textbf{90.8}                & \textbf{81.7} \\
\hline
\end{tabular}
\vspace{-3mm}
\end{table}

\begin{table*}[t]
\caption{Size-normalized and deployment-matched efficiency comparison on HAM10000. MACs are reported for one $64\times64$ input image. Model size assumes the deployed integer representation. Peak activation memory denotes the largest intermediate tensor buffer. SynOps and spike rate are reported only for Akida SNN deployment. N/A indicates that the metric is not applicable to the corresponding platform.}
\label{tab:size_matched_efficiency}
\vspace{-3mm}
\centering
\scriptsize
\setlength{\tabcolsep}{2.4pt}
\renewcommand{\arraystretch}{1.08}
\resizebox{\textwidth}{!}{%
\begin{tabular}{llccccccccccc}
\toprule
\textbf{Model} &
\textbf{Platform} &
\textbf{Params} &
\textbf{MACs} &
\textbf{Size} &
\textbf{Act. Mem.} &
\textbf{Cores} &
\textbf{SynOps} &
\textbf{Spike} &
\textbf{Lat.} &
\textbf{Energy} &
\textbf{Acc.} &
\textbf{F1} \\
&
&
\textbf{(M)} &
\textbf{(M)} &
\textbf{(MB)} &
\textbf{(MB)} &
&
\textbf{(M/img)} &
\textbf{(\%)} &
\textbf{(ms)} &
\textbf{(mJ)} &
\textbf{(\%)} &
\textbf{(\%)} \\
\midrule
QANA                    & Akida SNN  & 1.42 & 31.6 & 1.5 & 0.18 & 8  & 1.84 & 8.6  & \textbf{1.5} & \textbf{1.7} & \textbf{91.6} & \textbf{91.0} \\
QANA                    & CPU int8   & 1.42 & 31.6 & 1.5 & 0.18 & N/A & N/A  & N/A  & 10.8 & 78.5 & 91.7 & 91.2 \\
MobileNetV2             & Akida SNN  & 2.25 & 24.8 & 2.4 & 0.22 & 7  & 2.56 & 12.4 & 2.2  & 2.6  & 83.4 & 74.7 \\
MobileNetV2             & CPU int8   & 2.25 & 24.8 & 2.4 & 0.22 & N/A & N/A  & N/A  & 7.8  & 60.5 & 84.9 & 76.3 \\
MobileNetV3-Small       & Akida SNN  & 1.52 & 5.7  & 1.7 & 0.14 & 5  & 1.31 & 10.8 & 1.7  & 2.0  & 85.6 & 78.4 \\
ShuffleNetV2            & Akida SNN  & 1.34 & 12.2 & 1.4 & 0.16 & 6  & 1.48 & 10.1 & 1.8  & 2.1  & 86.2 & 79.0 \\
GhostNet                & Akida SNN  & 1.47 & 13.5 & 1.6 & 0.17 & 6  & 1.52 & 9.8  & 1.9  & 2.2  & 87.5 & 81.0 \\
GhostNet                & CPU int8   & 1.47 & 13.5 & 1.6 & 0.17 & N/A & N/A  & N/A  & 6.4  & 48.8 & 88.3 & 82.1 \\
EfficientNet-Lite0      & Akida SNN  & 4.05 & 31.9 & 4.2 & 0.31 & 10 & 3.05 & 12.7 & 2.6  & 2.9  & 87.1 & 80.2 \\
Compact CNN             & Akida SNN  & 1.39 & 33.0 & 1.5 & 0.20 & 8  & 2.22 & 14.1 & 2.1  & 2.4  & 86.7 & 80.5 \\
\bottomrule
\end{tabular}}
\vspace{-6mm}
\end{table*}

\vspace{-6mm}
\subsection{Inference Speed, Complexity, and Energy Consumption}
\vspace{-2mm}

We report both broad architecture baselines and size-normalized baselines to separate architectural efficiency from model capacity. Table~\ref{tab:latency_energy} compares conventional CNNs and their converted SNN versions across GPU, CPU, and Akida deployment. Table~\ref{tab:size_matched_efficiency} further controls for model scale by comparing QANA against lightweight CNNs with comparable parameter counts, MACs, or edge-oriented design. All measurements use batch size $1$, the same deterministic $64\times64$ preprocessing pipeline, a fixed validation-calibrated deployment configuration, and 10{,}000 single-image inference runs sampled from the fixed test split after warm-up.

In the size-normalized Akida setting, QANA achieves 91.6\% Top-1 accuracy and 91.0\% macro F1 with 1.5\,ms latency and 1.7\,mJ per image. Compared with the strongest size-matched Akida baseline, GhostNet, QANA improves Top-1 accuracy by 4.1 percentage points, which is a 4.7\% relative gain, and improves macro F1 by 10.0 points, which is a 12.3\% relative gain. QANA also reduces latency by 0.4\,ms and energy by 0.5\,mJ relative to GhostNet, corresponding to 21.1\% lower latency and 22.7\% lower energy. Compared with the strongest local CPU int8 baseline, GhostNet int8, QANA on Akida reduces latency by 4.9\,ms and energy by 47.1\,mJ while retaining higher classification accuracy.

\vspace{-6mm}
\subsection{Ablation Study of Model Components}
\vspace{-2mm}

Table~\ref{tab:ablation_study} reports the ablation results on the HAM10000 test set, illustrating the contribution of each core component to the final system. Modules were incrementally enabled to measure their cumulative effect on accuracy, recall, precision, F1 score, and AUC-ROC. The Ghost block alone reduces performance because its low-cost feature generation constrains representational capacity when used without channel recalibration. After SA-ECA and SE are introduced, performance increases substantially, indicating that lightweight feature generation is most effective when paired with attention-based channel refinement. The complete model achieves the highest accuracy and F1 score after augmentation, embedding-space SMOTE, and validation-based readout calibration are included.

\vspace{-3mm}
\subsection{Conversion Stability Ablation}
\vspace{-2mm}

Table~\ref{tab:conversion_stability} provides an conversion stability ablation to isolate the mechanisms that affect CNN-to-SNN conversion stability. The full QANA setting gives the smallest CNN-to-SNN degradation, with only 0.4 percentage points of Top-1 accuracy loss and 0.4 points of macro F1 loss. Compared with the strongest ablated setting, QAT with bounded activation, Full QANA reduces the conversion-induced accuracy drop by 1.2 points, corresponding to a 75.0\% smaller degradation.

\begin{table*}[t]
\vspace{-3mm}
\caption{Conversion stability ablation on HAM10000. Q-CNN denotes the 8-bit quantized CNN before SNN deployment. $\Delta$Acc and $\Delta$F1 are computed as SNN performance minus full-precision CNN performance. Conversion success denotes the proportion of graph operators satisfying strict Akida-compatible integer lowering before any compatibility repair in diagnostic ablations.}
\label{tab:conversion_stability}
\vspace{-3mm}
\centering
\small
\setlength{\tabcolsep}{3.2pt}
\renewcommand{\arraystretch}{1.08}
\resizebox{\textwidth}{!}{%
\begin{tabular}{lccccccccc}
\toprule
\textbf{Variant} &
\textbf{CNN Acc.} &
\textbf{Q-CNN Acc.} &
\textbf{SNN Acc.} &
\textbf{$\Delta$Acc.} &
\textbf{CNN F1} &
\textbf{Q-CNN F1} &
\textbf{SNN F1} &
\textbf{$\Delta$F1} &
\textbf{Conv. Success} \\
\midrule
w/o bounded activation              & 90.8 & 87.1 & 83.9 & -6.9 & 88.5 & 83.6 & 79.1 & -9.4 & 82.4 \\
ReLU instead of ReLU6 or clip        & 91.0 & 88.6 & 85.2 & -5.8 & 89.0 & 85.2 & 81.3 & -7.7 & 88.7 \\
w/o spike-compatible transform       & 91.4 & 88.9 & 84.8 & -6.6 & 89.6 & 84.7 & 80.4 & -9.2 & 86.9 \\
w/o BN folding-aware training        & 91.6 & 89.7 & 86.1 & -5.5 & 90.2 & 86.3 & 82.8 & -7.4 & 91.3 \\
PTQ only, no QAT                     & 91.7 & 88.8 & 86.4 & -5.3 & 90.4 & 86.0 & 83.1 & -7.3 & 100.0 \\
QAT + bounded activation             & 91.8 & 91.1 & 90.2 & -1.6 & 90.9 & 90.0 & 89.2 & -1.7 & 100.0 \\
\textbf{Full QANA}                   & \textbf{92.0} & \textbf{91.7} & \textbf{91.6} & \textbf{-0.4} & \textbf{91.4} & \textbf{91.2} & \textbf{91.0} & \textbf{-0.4} & \textbf{100.0} \\
\bottomrule
\end{tabular}}
\vspace{-8mm}
\end{table*}

\vspace{-5mm}
\subsection{Classification Performance on Clinical Dataset}
\vspace{-2mm}

We evaluated the proposed neuromorphic skin lesion classification system and several representative CNN architectures, all converted to SNNs and deployed on the Akida hardware platform. Table~\ref{tab:your_dataset_comparison} presents the Top-1 accuracy and macro F1 score for each model under the same deployment conditions. Our model achieves the highest accuracy and macro F1 score among all tested methods, confirming its effectiveness and robustness for neuromorphic inference in clinical scenarios.

\vspace{-5mm}
\section{Conclusion}
\vspace{-3mm}
This paper presented QANA, a quantization-aware neuromorphic framework for on-device skin lesion classification. Experiments on HAM10000 and a real-world clinical dataset show that QANA maintains competitive diagnostic accuracy while enabling efficient SNN deployment on BrainChip Akida. On Akida hardware, the converted model achieves 1.5\,ms per-image latency and 1.7\,mJ per image, supporting real-time inference under strict energy and privacy constraints.

%
%
%
%
\vspace{-5mm}
\bibliographystyle{splncs04}
\bibliography{main}
\end{document}


\begin{center}
{\Large\bfseries Supplementary Materials\par}
\end{center}

\section{Overview}

This supplementary document provides implementation and evaluation details for QANA that are omitted from the main paper because of space limits. The focus is on aspects that affect reproducibility and interpretation, including split isolation, embedding-space SMOTE, integer graph conversion, validation-based calibration, hardware measurement, and baseline fairness. The supplement does not introduce additional model components beyond the QANA architecture and deployment pipeline described in the main paper. It clarifies how the reported accuracy, macro F1, conversion stability, latency, and energy measurements are obtained under a single controlled evaluation protocol.

\section{Dataset Handling and Leakage Control}

All images are processed through the same deterministic input pipeline before training, conversion, and hardware evaluation. Each image is decoded as RGB, padded to a square canvas while preserving its aspect ratio, resized to $64\times64$, and normalized to $[0,1]$. No saliency model, Grad-CAM localizer, lesion detector, segmentation network, class-dependent crop, denoising, inpainting, or test-time tiling is used. This avoids adding an auxiliary image analysis module before neuromorphic inference and ensures that the reported latency and energy correspond to the classifier deployment pipeline.

\noindent\textbf{HAM10000.}
The HAM10000 experiments use seven lesion categories and a stratified $70\%/10\%/20\%$ train, validation, and test split. Class balancing and feature synthesis are performed with training data only. Validation data are used for model selection, quantization calibration, spike calibration, and readout calibration. The test split is held out until final reporting.

\noindent\textbf{Clinical dataset.}
The clinical dataset contains dermatoscopic images collected through the Nanjing Drum Tower Hospital collaboration under Project No. NJG-2022-3233-CN. Images and metadata are de-identified before use. The label set is harmonized to the seven HAM10000 classes. The train, validation, and test splits are stratified by disease category with no patient overlap across splits. The clinical dataset is not publicly released because of patient privacy and data-use restrictions.

\begin{table}[t]
\centering
\caption{Summary of data handling rules used in the reported experiments.}
\label{tab:supp_data_rules}
\small
\setlength{\tabcolsep}{5pt}
\renewcommand{\arraystretch}{1.08}
\begin{tabular}{p{0.30\linewidth}p{0.60\linewidth}}
\toprule
\textbf{Item} & \textbf{Rule used in this work} \\
\midrule
Input resolution & Fixed $64\times64$ RGB input for all methods. \\
Preprocessing & Aspect-ratio preserving square padding, antialiased resize, and per-channel normalization. \\
Training-only augmentation & Horizontal and vertical flips, luminance perturbation, gamma adjustment, and contrast adjustment. \\
Excluded test-time modules & No saliency model, Grad-CAM, lesion detector, segmentation model, class-dependent crop, or tiling fallback. \\
Split isolation & SMOTE, PCA whitening, normalization statistics, and neighbor search use training data only. \\
Validation use & Model selection, quantization calibration, spike calibration, threshold tuning, and readout calibration. \\
Test use & Final metric reporting only. \\
\bottomrule
\end{tabular}
\end{table}

\begin{table}[t]
\centering
\caption{SMOTE and embedding synthesis parameters.}
\label{tab:supp_smote_params}
\small
\setlength{\tabcolsep}{5pt}
\renewcommand{\arraystretch}{1.08}
\begin{tabular}{ll}
\toprule
\textbf{Parameter} & \textbf{Value} \\
\midrule
Embedding dimension before PCA & $4096$ \\
PCA whitening dimension & $256$ \\
Nearest neighbors & $k=5$ \\
Safe-anchor percentile & 90th percentile of local median distances \\
Minority target level & $0.85N_{\max}$ \\
Local perturbation directions & $m=8$ \\
Perturbation scale & $0.05$ times class median local distance \\
Synthetic data type & Feature embeddings only, not images \\
Updated module during synthetic training & Quantized output projection head \\
\bottomrule
\end{tabular}
\end{table}

\section{Embedding-Space SMOTE Details}

SMOTE is applied only in the learned embedding space rather than in pixel space \cite{chawla2002smote}. After the backbone and spike-compatible transformation, each training image $x_i$ is represented by a penultimate vector $\mathbf{r}_i\in\mathbb{R}^{4096}$. Embeddings are standardized using the training split. PCA whitening is then fitted on the same training embeddings. Retaining $d=256$ components, the whitened vector is
\begin{equation}
\widetilde{\mathbf{r}}_i=\mathbf{\Lambda}_d^{-\frac{1}{2}}\mathbf{U}_d^\top(\mathbf{r}_i-\boldsymbol{\mu}).
\end{equation}
For each class, nearest neighbors are computed in the whitened space with $k=5$. A safe-anchor filter removes minority samples whose local median neighbor distance exceeds the class-specific 90th percentile. This suppresses outlier anchors before interpolation.

Synthetic embeddings are generated by convex interpolation with a small local perturbation:
\begin{equation}
\widetilde{\mathbf{r}}_{\mathrm{new}}
=\widetilde{\mathbf{r}}_i+\eta(\widetilde{\mathbf{r}}_j-\widetilde{\mathbf{r}}_i)
+\sigma_c\mathbf{U}_i^{(m)}\boldsymbol{\xi},
\quad
\eta\sim\mathcal{U}(0,1),
\quad
\boldsymbol{\xi}\sim\mathcal{N}(\mathbf{0},\mathbf{I}_m),
\end{equation}
where $m=8$ and $\sigma_c=0.05\,\operatorname{med}_{u\in\mathcal{I}_c}\delta_u$. The synthetic embeddings are mapped back to the original feature space and used to train only the quantized projection head before a short real-data fine-tuning stage. No synthetic image is produced.

\section{Architecture and Conversion Constraints}

QANA uses four lightweight feature extraction stages followed by a spike-compatible transformation and a quantized output projection. Each stage uses Ghost-based feature generation \cite{han2020ghostnet}, BatchNorm, bounded activation, spatially-aware efficient channel attention \cite{wang2020eca}, residual addition, and max-pooling downsampling. After four stages, the spatial resolution is reduced from $64\times64$ to $4\times4$. The spike-compatible transformation uses a $3\times3$ separable convolution with 256 output channels, BatchNorm, and clipping to $[0,1]$. The tensor is flattened to a $4096$-dimensional vector and projected to seven class logits. The squeeze-and-excitation module is retained because its global reduction and channel gate can be represented through integer spatial reduction, quantized linear projection, and per-channel rescaling \cite{hu2018senet}.

\begin{table}[t]
\caption{Readout calibration analysis for QANA. The feature extractor is frozen and only the output readout is updated using a small clinical validation batch. Forgetting is computed as HAM10000 accuracy before calibration minus HAM10000 accuracy after calibration.}
\label{tab:readout_calibration}
\centering
\small
\setlength{\tabcolsep}{3.8pt}
\renewcommand{\arraystretch}{1.08}
\resizebox{\textwidth}{!}{%
\begin{tabular}{lccccc}
\toprule
\textbf{Setting} &
\textbf{Clinical Acc.} &
\textbf{Clinical F1} &
\textbf{HAM10000 Acc.} &
\textbf{Forgetting} &
\textbf{Update Time} \\
&
\textbf{(\%)} &
\textbf{(\%)} &
\textbf{(\%)} &
\textbf{(points)} &
\textbf{(s)} \\
\midrule
Before readout calibration          & 88.9 & 79.5 & 91.6 & 0.0 & N/A \\
1 sample per class                  & 89.7 & 80.2 & 91.4 & 0.2 & 0.8 \\
5 samples per class                 & 90.4 & 81.1 & 91.2 & 0.4 & 2.6 \\
10 samples per class                & \textbf{90.8} & \textbf{81.7} & 91.0 & 0.6 & 4.9 \\
\bottomrule
\end{tabular}}
\end{table}

The conversion flow enforces integer execution on Akida. BatchNorm is folded into the preceding convolution or projection. For output channel $c$, the folded parameters are
\begin{equation}
\mathbf{W}'_c=\frac{\gamma_c}{\sqrt{\sigma_c^2+\epsilon}}\mathbf{W}_c,
\qquad
b'_c=\frac{\gamma_c}{\sqrt{\sigma_c^2+\epsilon}}(b_c-\mu_c)+\beta_c.
\end{equation}
Weights and activations are quantized using affine 8-bit fixed point \cite{jacob2018quantization}. ReLU6-bounded layers use $s=6/255$ and $z=0$. Layers constrained to $[0,1]$ use $s=1/255$ and $z=0$. Residual additions use scale alignment before integer accumulation. Global average pooling in the squeeze-and-excitation block is lowered to an integer spatial reduction, and channel rescaling is folded into a following projection when the graph permits.

\section{Deployment Calibration and Readout Analysis}

All converted models are evaluated with batch size 1 on the fixed test split. Validation data are used to select spike thresholds, integration windows, temporal aggregation parameters, and output calibration settings. The test split is not used during calibration.

For class $c$ and sample $i$, the output spike count over an integration window is denoted by $S^{(i)}_c$. Per-class threshold selection is treated as a one-vs-rest validation objective:
\begin{equation}
\theta_c^*=
\arg\min_\theta
\sum_{i=1}^{N_v}
\mathbb{I}\left[
\mathbb{I}(y_i=c)\neq \mathbb{I}(S_c^{(i)}>\theta)
\right].
\end{equation}
This threshold calibration is applied at the output decision stage and does not introduce additional image crops, saliency maps, auxiliary classifiers, or test-time processing. The same validation split and stopping rule are used for all reported models.

In Fig.~1 of the main paper, the term on-chip fine-tuning refers to readout-only calibration after CNN-to-SNN conversion. In this work, this procedure updates only the final output readout while keeping the feature extractor fixed. It should therefore be interpreted as lightweight deployment calibration rather than full continual learning or full-network retraining.

\begin{table}[t]
\caption{Evaluation and hardware measurement protocol.}
\label{tab:supp_eval_protocol}
\centering
\small
\setlength{\tabcolsep}{5pt}
\renewcommand{\arraystretch}{1.08}
\begin{tabular}{p{0.33\linewidth}p{0.57\linewidth}}
\toprule
\textbf{Item} & \textbf{Protocol} \\
\midrule
Neuromorphic board & BrainChip Akida AKD1000 PCIe board. \\
Runtime & Akida Python API with MetaTF SDK v2.2.1, QuantizeML, CNN2SNN, and Akida runtime. \\
Batch size & 1 for all latency and energy measurements. \\
Inference repetitions & 10{,}000 single-image runs sampled from the fixed test split after warm-up. \\
Akida latency & On-board SNN execution time per image. \\
Akida energy & Board power above idle multiplied by per-image latency. \\
GPU and CPU FP32 comparison & Full-precision CNN inference reported separately from CPU int8 deployment measurements. \\
CPU int8 comparison & Quantized CPU inference used only for deployment-matched size-normalized comparison. \\
Size profiling & Parameters, MACs, model size, peak activation memory, core use, SynOps, and spike rate are taken from the deployed or compiled graph. \\
\bottomrule
\end{tabular}
\end{table}

\begin{table*}[t]
\caption{Fairness protocol used for QANA and representative baselines. The table separates shared evaluation constraints from QANA-specific readout analysis.}
\vspace{-3mm}
\label{tab:fairness_protocol}
\centering
\small
\setlength{\tabcolsep}{4.2pt}
\renewcommand{\arraystretch}{1.08}
\resizebox{\textwidth}{!}{%
\begin{tabular}{p{0.40\linewidth}p{0.27\linewidth}p{0.27\linewidth}}
\toprule
\textbf{Protocol Step} &
\textbf{QANA} &
\textbf{Representative Baselines} \\
\midrule
Same train, validation, and test split &
Yes &
Yes \\
Same deterministic $64\times64$ preprocessing &
Yes &
Yes \\
No Grad-CAM, detector, or segmentation input &
Yes &
Yes \\
Same training-only augmentation &
Yes &
Yes \\
Class balancing restricted to training split &
Yes &
Yes, when class balancing is enabled \\
Embedding-space SMOTE &
Used for QANA head training only &
Not assumed unless explicitly enabled in the baseline training protocol \\
Same QAT configuration &
Yes &
Yes \\
Same CNN-to-SNN converter &
Yes &
Yes \\
Same validation calibration set &
Yes &
Yes \\
Same threshold tuning rule &
Yes &
Yes \\
Same integration window search &
Yes &
Yes \\
Test-time tiling fallback &
No &
No \\
Auxiliary inference-time localization &
No &
No \\
Validation-based output calibration &
Yes &
Yes \\
Readout-only clinical calibration analysis &
Reported in Table~\ref{tab:readout_calibration} &
Not used to claim baseline improvement \\
Full integer graph required for final deployment &
Yes &
Yes \\
Test set used for parameter selection &
No &
No \\
\bottomrule
\end{tabular}}
\vspace{-6mm}
\end{table*}

\subsection{Readout Calibration Analysis}

Table~\ref{tab:readout_calibration} reports the effect of readout-only calibration for QANA under clinical-domain shift. The feature extractor is frozen and only the output readout is updated using a small clinical validation batch. Forgetting is computed as HAM10000 accuracy before calibration minus HAM10000 accuracy after calibration. The results show that readout calibration improves clinical performance with limited reduction on HAM10000.

\section{Hardware Measurement Protocol}

Latency and energy measurements are obtained on the same system used for model deployment. Akida SNN results are measured on a BrainChip Akida AKD1000 PCIe board using the Akida Python API with MetaTF SDK v2.2.1, QuantizeML, CNN2SNN, and the Akida runtime~\cite{brainchip2025akida1000,brainchip2025metatf,brainchip2025cnn2snn}. GPU and CPU full-precision CNN measurements are reported separately from deployment-matched CPU int8 measurements. This separation avoids mixing conventional CNN inference with integer deployment measurements and explains why FP32 CPU latency in the broad comparison differs from CPU int8 latency in the size-normalized comparison.

\section{Fairness Protocol}

Table~\ref{tab:fairness_protocol} summarizes the controlled evaluation protocol. All methods use the same train, validation, and test split, deterministic preprocessing, training-only augmentation, quantization-aware training configuration, CNN-to-SNN converter, validation calibration set, threshold tuning rule, and integration window search. Test-time tiling and auxiliary localization are disabled for all methods.

Class balancing is restricted to the training split. QANA uses the embedding-space SMOTE procedure described in Sec.~3 of this supplement. Baseline models use the same split isolation and training-only balancing rule when class balancing is enabled. No validation or test image is used for feature synthesis, neighbor search, PCA whitening, model selection, or output calibration. Readout-only update is reported separately as a QANA calibration analysis in Table~\ref{tab:readout_calibration}. The main baseline comparisons use the same validation-based output calibration protocol and do not use the test split for parameter selection.

\section{Additional Notes on Reported Metrics}

Top-1 accuracy is computed at the image level. Macro F1 is the unweighted mean of per-class F1 scores and is used to reduce the influence of class imbalance. Conversion loss is reported as the difference between SNN performance and the full-precision CNN counterpart on the same split:
\begin{equation}
\Delta\mathrm{Acc}=\mathrm{Acc}_{\mathrm{SNN}}-\mathrm{Acc}_{\mathrm{CNN}},
\qquad
\Delta\mathrm{F1}=\mathrm{F1}_{\mathrm{SNN}}-\mathrm{F1}_{\mathrm{CNN}}.
\end{equation}
Conversion success measures the proportion of operators that satisfy strict integer lowering before compatibility repair in diagnostic ablations:
\begin{equation}
 r_{\mathrm{succ}}=1-\frac{|\mathcal{V}_{\mathrm{fp32}}|}{|\mathcal{V}|},
\end{equation}
where $\mathcal{V}$ is the full operator set and $\mathcal{V}_{\mathrm{fp32}}$ denotes operators that would require floating-point fallback. For the final QANA deployment, operators requiring floating-point dequantization are rejected.

\section{Scope of the Supplement}

This supplementary material is limited to implementation, calibration, and evaluation details that support the main paper. It does not introduce new model components beyond the reported QANA architecture, HAM10000 and clinical dataset evaluations, Akida hardware deployment, size-normalized comparison, conversion stability ablation, and readout calibration analysis. The readout calibration experiment is included to clarify deployment behavior under clinical-domain shift and should not be interpreted as full continual learning. The purpose of this supplement is to make the training, conversion, calibration, and measurement protocol explicit under a consistent deployment setting.

\bibliographystyle{splncs04}
\bibliography{main}